\documentclass[aps,pra,numerical,superscriptaddress,showpacs,
reprint
]{revtex4-1}
\usepackage{amsmath}
\usepackage{color}
\usepackage{graphicx}
\usepackage[english]{babel}
\usepackage[noload=abbr]{siunitx}

\begin{document}
\title{High vacuum compatible fiber feedthrough for hot alkali vapor cells}

\author{Daniel Weller}
\affiliation{5.~Physikalisches Institut and Center for Integrated Quantum Science and Technology, University of Stuttgart, Pfaffenwaldring 57, 70569 Stuttgart, Germany}
\author{Arzu Yilmaz}
\affiliation{Physikalisches Institut, Eberhard Karls University, T\"ubingen, Germany}
\author{Harald K\"ubler}
\affiliation{5.~Physikalisches Institut and Center for Integrated Quantum Science and Technology, University of Stuttgart, Pfaffenwaldring 57, 70569 Stuttgart, Germany}
\author{Robert L\"ow}
\email{r.loew@physik.uni-stuttgart.de}
\affiliation{5.~Physikalisches Institut and Center for Integrated Quantum Science and Technology, University of Stuttgart, Pfaffenwaldring 57, 70569 Stuttgart, Germany}

\date{\today}
  
\begin{abstract}
We demonstrate a method to produce vacuum tight, 
metal free,
bakeable and alkali compatible feedthroughs for optical fibers.
The manufacturing process mainly relies on 
encasing fibers made of fused silica with glass materials with lower melting points by heating.
We analyze the vacuum and optical performance of the devices
by various methods including helium leak checking and several spectroscopic schemes,
among others electromagnetically induced transparency involving Rydberg states.
\end{abstract}

\maketitle

\section{Introduction}
The spectroscopy of ultracold atomic gases,
ions and molecules as well as 
atomic and molecular vapors at room temperature
typically requires vacuum conditions ranging from
\SIrange[parse-numbers = false]{10^{-11}}{10^{-3}}{\milli\bar}.
For an optimal matter light coupling,
it is desirable to directly apply optical fibers 
close to the atomic sample
which requires suitable vacuum feedthroughs.
For stainless steel vacuum chambers with KF or CF systems
such feedthroughs exist, 
based on PTFE and metallic ferrules in a swage-lock type configuration \citep{abraham1998teflon,miller2001all}. 
They are used for example to setup fiber-pigtailed optical tweezers for single-atom trapping \cite{Garcia2013fiber},
or to insert tapered fibers
into clouds of laser-cooled atoms \cite{le2015electromagnetically}.
A more simple approach is to just seal the feedthrough with epoxy,
e.g.\,used to install microscopic Fabry-P\'erot cavities \cite{colombe2007strong}.
The drawback of such feedthroughs is their 
limited vacuum compatibility,
especially when exposed to highly reactive and corrosive alkali vapors,
which results in chemical reactions and additional gas loads.
The consequence is the need for permanent pumping to maintain the vacuum.
Another point is a reduced thermal resistance,
which only allows bake-out temperatures of up to \SI{120}{\celsius}.
An alternative method is to couple the light through a chamber window
into the fiber inside the vacuum \cite{epple2014rydberg}
for the price of reduced coupling efficiency and stability.
For enclosed vacuum systems without a pump,
as it is standard for spectroscopy cells made of glass,
it is mandatory to have an absolute vacuum tight feedthrough made of materials with low outgassing rates.
One alternative is to fill hollow core fibers under vacuum conditions with a gas
and then to seal the ends vacuum tightly \cite{benabid2005compact}.
For a more general applicability,
it is desirable to have a versatile fiber feedthrough only made of glass materials
to avoid out-gassing, chemical reactions and coupling losses
while still having the ability to heat the system to high temperatures. 
In the following,
we explain the production and test of such feedthroughs,
and conclude with applications made possible by this technique.

\section{Manufacturing Process} \label{sec:manufacturing}
The manufacturing process relies on the diverging softening points of different glass materials.
Here we combine glass types with a rather low melting point 
(borosilicate, \SI{825}{\celsius})
to vacuum tightly enclose an optical fiber with a much higher temperature stability
(fused silica, \SI{1665}{\celsius}).
The anticipation was
that the almost melted borosilicate adheres onto the fused silica fiber
and does not crack due to different expansion coefficients
(\SI{3.3e-6}{} vs.\,\SI{0.6e-6}{})
when temperatures are changed,
e.g.\,for a typical bake-out procedure.

Through a carefully performed manufacturing process as described below,
we have been able to produce such a heterogeneous combination of materials,
which also maintains the optical performance of the fibers.
For the purpose of a proof of principle,
we used standard optical single-mode fibers
(SMF, Thorlabs 780HP)
and multi-mode fibers
(MMF, Thorlabs FG050LGA).
\subsection{Production Steps}
\begin{figure}[]
\centering
\includegraphics[width=\linewidth]{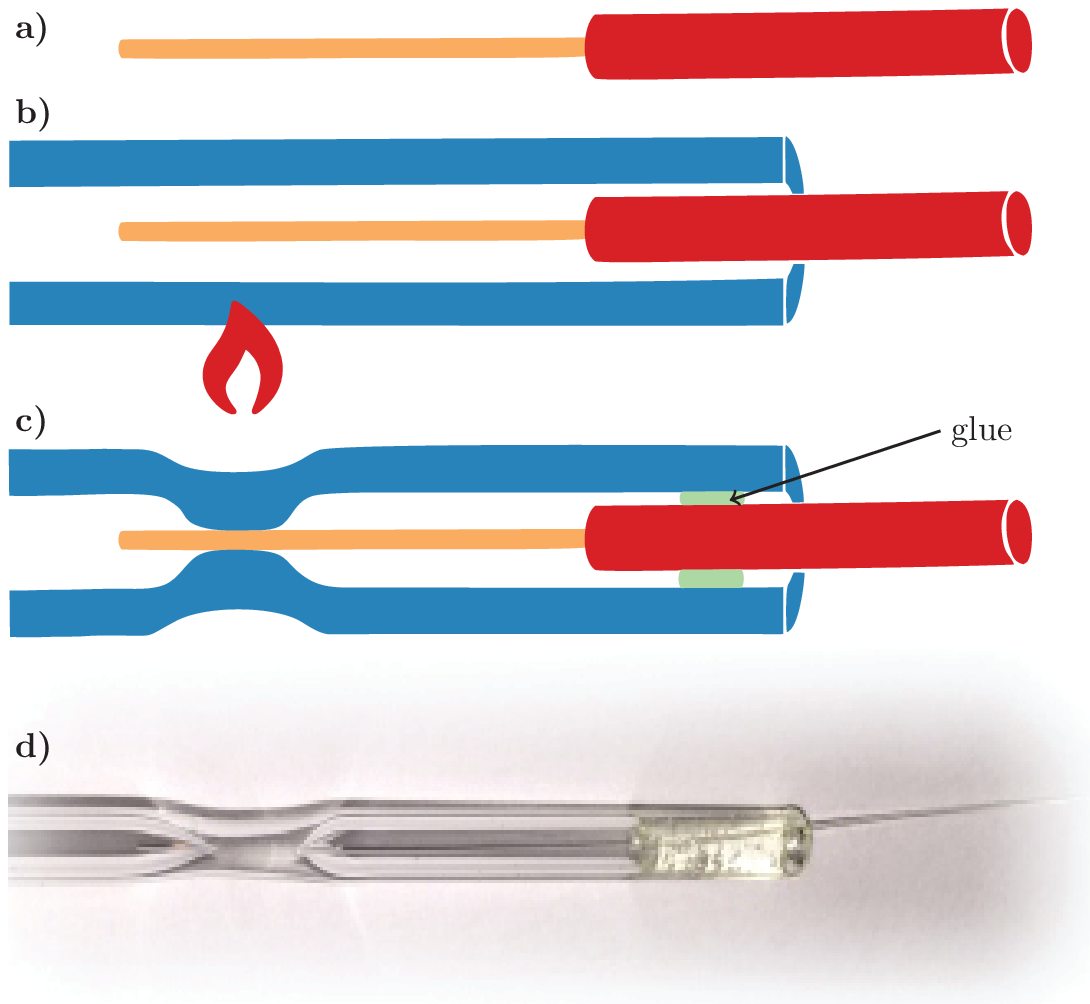}
\caption{Manufacturing Process:
\textbf{(a)} strip, cleave and clean fiber,
\textbf{(b)} insert fiber into tube with only slightly larger diameter (shrink beforehand if necessary) and uniformly heat e.g.\,by rotation,
\textbf{(c)} protect/fix the exposed end with suitable epoxy/glue
\textbf{(d)} actual photograph of a device.}
\label{fig:productionsteps}
\end{figure}

Before processing, we have mechanically stripped a SMF and a MMF on a length of about
\SI{3}{\centi\meter} (Fig.\,\ref{fig:productionsteps}(a))
and subsequently removed the residual coating by wiping with isopropyl alcohol soaked anti-static tissues to prevent dust contamination.
As a jacket we chose a borosilicate tube with 
\SI{2}{\milli\meter} inner (\SI{3}{\milli\meter} outer) diameter
which is capable of surrounding most commercial fibers.

To allow an optimal heating from all sides,
we installed the capillary inside a self-built lathe.
The lathe consists of two rotatable drill chucks
driven by synchronized stepper motors.
The distance between the mounts can be adjusted to a maximum length of
\SI{30}{\centi\meter}.
To reduce avoidable heating of the fibers,
we pre-shrink the inner diameter of the tube,
before inserting the fiber.
This is achieved by moderately heating the glass with a hand torch
(e.g.\,Proxxon Microflam)
until the inner diameter roughly matches the fiber cladding size.
When heated carefully, the glass-tube slowly starts to collapse by itself
due to the surface tension.
Both utilizing a lathe and pre-shrinking is no mandatory step,
but significantly facilitates the following process.

The fiber is then inserted into the tube.
Alignment is carried out such that the protective polyimide coating has enough distance 
($\approx$\SI{15}{\milli\meter})
to the sealing position,
but still remains within the tube.
The length of the stripped part of the fiber has to be adjusted accordingly beforehand.

The lathe is operated at around 3-5\,rps while heating the borosilicate tube with the hand torch 
(Fig.\,\ref{fig:productionsteps}(b)).
To avoid twisting the fiber,
either its length needs to be short enough,
or the rest of the fiber is coiled up and fixed onto the rotating parts of the lathe.
As the tube collapses onto the fiber,
the torch can be moved along the axis to further increase the length of the seal.
Due to the limited heat of the hand torch and the short exposure (around \SI{10}{\second}),
the melting point of the fused silica is not reached,
preserving both functionality and shape of the fiber.
However, the melting point of the borosilicate is exceeded. 
Due to surface tension, the tube collapses and forms a durable
and vacuum tight seal around the fiber cladding (Fig.\,\ref{fig:faces}).

The connection between tube and fiber is distinctly brittle
because of the lacking cladding.
Especially slight twists are likely to break the device.
To prevent such accidental rupture,
a suitable glue is then applied to the coated part of the fiber
inside the tube but outside of the intended vacuum part.
(Fig.\,\ref{fig:productionsteps}(c)).
As the glue only provides mechanical stability,
the choice here is not critical,
as long as temperature compatibility is given.
We used Epo-Tek 301.
An exemplary photograph of such a feedthrough is shown in 
Fig.\,\ref{fig:productionsteps}(d).

The finalized feedthrough can be attached to more elaborate devices,
such as the flat polished window of a precision spectroscopy cell.
Alternatively, the tube is integrated into a construction in advance
and the sealing process is carried out subsequently.

\subsection{Sample Preparation}
To later perform spectroscopic analysis of the devices,
the assembled feedthroughs were attached to a larger tube
with a flattened region for free beam spectroscopy.
The devices were evacuated to \SI{1e-7}{\milli\bar},
each filled with a droplet of rubidium
and sealed conventionally on the fiber-averted side.
The vapor pressure has been adjusted by temperature.

\section{Performance Analysis}
\subsection{Vacuum Performance}
For helium leak checks, 
the devices were attached to an all-glass KF16 flange and
connected to a commercial leak detector (Leybold Phoenixl 300).
The helium leak-rate was below the detection limit of \SI{1e-12}{\milli\bar \liter \per \second},
measured at both room temperature and \SI{130}{\celsius}.
Apparently,
the difference in thermal expansion of borosilicate and fused silica
(\SI[per-mode=reciprocal]{3.3e-6}{\per\kelvin} and \SI[per-mode=reciprocal]{0.6e-6}{\per\kelvin}, respectively)
plays only a very minor role for the vacuum tightness:
For the typical diameter of a stripped fiber of \SI{125}{\micro\meter},
a temperature change of \SI{100}{\kelvin}
would yield a theoretical difference of diameters of around \SI{34}{\nano\meter}.
In a very simplified picture,
a pipe of diameter $d = \SI{34}{\nano\meter}$ and length $l=\SI{2}{\milli\meter}$ has a through-put of \cite{steckelmacher1966review}
\begin{equation}
q_{\mathrm{pV}} = \sqrt{\frac{2\pi k_\mathrm{B} T}{m}}\cdot\frac{d^3}{6 l}\Delta p ,
\label{eq:leakrate}
\end{equation}
with Boltzmann constant $k_\mathrm{B}$, temperature $T$, molecular mass $m$ and pressure gradient $\Delta p$.
For helium and nitrogen (N$_2$) at \SI{400}{\kelvin} and a pressure difference of \SI{1}{\bar},
the numerical values are
$q_{\mathrm{pV}} = 7.5\,\times 10^{-11} \SI{}{\milli\bar\liter\per\second}$ 
and $2.8\,\times 10^{-11} \SI{}{\milli\bar\liter\per\second}$, respectively.
Both values are
already above the measured rate.
The \textit{actual} gap should have a much larger cross section than the aforementioned tube
(area $10^4$ times larger).
Presumably, this gap is much smaller or even nonexistent,
for instance due to elasticity of the fiber.

\begin{figure}[htbp]
\centering
\includegraphics[width=\linewidth]{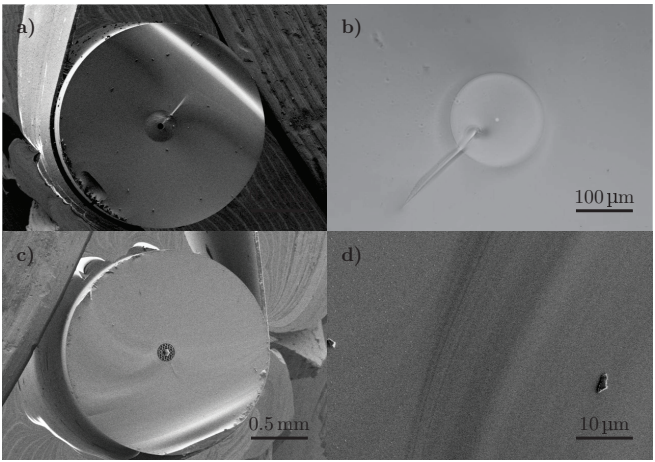}
\caption{Top view SEM images (a,c,d) and light microscopy image (b) of the cleaved facets of assemblies with different fiber types embedded. \textbf{(a)} \SI{60}{\micro\meter}-core capillary, the black circle is the hollow core, surrounded by the capillary cladding (diameter \SI{0.25}{\milli\meter}) and the larger tube). \textbf{(b)} Commercial SMF \textbf{780-HP}, the core is the bright spot, enclosed in cladding (\SI{125}{\micro\meter}) and tube. The diagonal feature is an artefact from cleaving. \textbf{(c)} Hollow-Core Photonic Crystal fiber. The characteristic structure is clearly visible, but the transition towards the tube shows no contrast. \textbf{(d)} Close-up of the interface between the fiber in (c) and the surrounding tube.
}
\label{fig:faces}
\end{figure}
To further substantiate the smoothness of the interface between fiber and capillary,
we examined the profile with scanning electron microscopy (Fig.\,\ref{fig:faces}).
For room temperature, apparently the tube fuses onto the fiber surface without a gap.

\subsection{Spectroscopic Pressure Observation}
A convenient way to study the vacuum condition inside the vapor cell
are narrow-band spectroscopy schemes,
which are sensitive to collisions with a potential background gas.
Such a perturbing gas can either result from a 
physical or a virtual leak,
from a chemical reaction of rubidium with residual contaminants
or from outgassing of the doped fiber.

\begin{figure}[htbp]
\centering
\includegraphics[width=\linewidth]{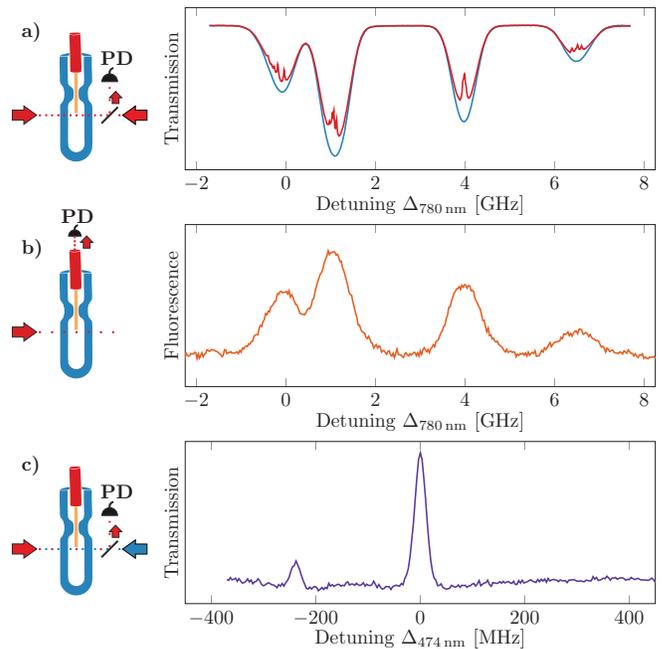}
\caption{Spectroscopic examples. \textbf{(a)} Absorption (blue) and saturation spectroscopy (red). \textbf{(b)} Fluorescence signal coupled out of the multi-mode fiber. \textbf{(c)} EIT signal using Rydberg state $22\mathrm{S}_{1/2}$, with hyperfine splitting in the intermediate state. In all schemes, the laser light was applied via free space propagating beams. The fiber was used to collect the fluorescence light in (b). The arrows (left) indicate the laser propagation, the signal is obtained by a photo-diode (PD).}
\label{fig:plots}
\end{figure}

The rubidium filled sample is heated
and kept at temperatures above \SI{100}{\celsius} for 30 days.
Different spectroscopy schemes
with different sensitivities to a background gas
are then applied repeatedly.
Fig.\,\ref{fig:plots} displays typical spectra captured for analysis.

Being easy to set up and widely available,
we first applied absorption and saturation spectroscopy.
Fig.\,\ref{fig:plots}(a) 
shows resulting measurements of both absorption (blue curve) 
and saturation spectroscopy (red curve) 
of the 
$5$S$_{1/2} \rightarrow 5$P$_{3/2}$ transition.
The Lamb-Dips are well resolved,
and thus motivate further analysis.
The fluorescence signal can be easily captured with the multi-mode fiber.
Fig.\,\ref{fig:plots}(b) shows the outcoupled light,
when exciting atoms in front of the cleaved fiber tip inside the cell.

A much more sensitive probe of the vacuum are highly excited Rydberg states,
which are just by their mere size more likely to collide
with atoms from a potential background gas.
To provide a quantitative measure for the background pressure
the probe line-width of the Electromagnetically Induced Transparency (EIT) ladder scheme
$5\mathrm{S}_{1/2}\, \mathrm{F}=3 \rightarrow 5\mathrm{P}_{1/2}\, \mathrm{F}=4 \rightarrow 22\mathrm{S}_{1/2}\, \mathrm{F'}=3$
is monitored (Fig.\,\ref{fig:plots}(c)).
No significant increase in line-width is observed:
the FWHM deviated less than \SI{5}{\mega\hertz} from the initial measure of
\SI{13}{\mega\hertz},
and remained below \SI{15}{\mega\hertz} at the end of the observation period.
As EIT spectroscopy was possible during the whole period,
the background pressure can be assumed to be stable.
Assuming nitrogen as the disturber gas (i.e.\,\textit{leaked air}),
this would correspond to a final absolute pressure of \SI{0.1}{\milli\bar}, 
or a (constant) leak rate of $Q=\SI{1e-10}{\milli\bar\liter\per\second}$ over the whole period of 30 days
\cite{petitjean1984thermal}.

\subsection{Optical Properties}
\begin{figure}[htbp]
\centering
\includegraphics[width=\linewidth]{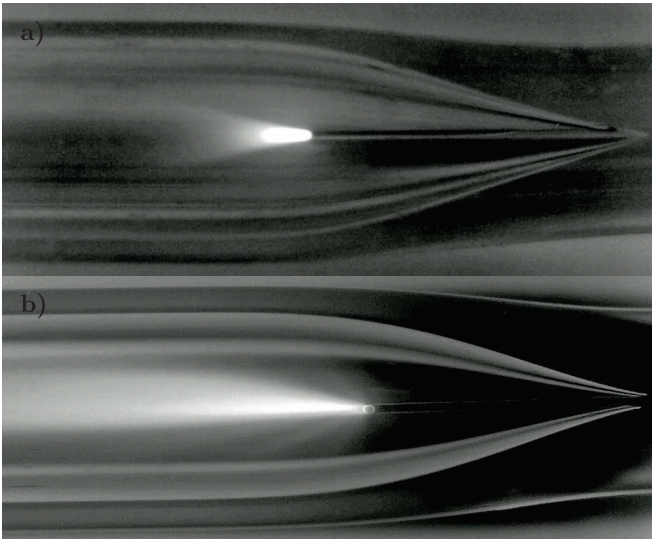}
\caption{Fluorescence cone of the outcoupled light in rubidium vapor.
 \textbf{a)} Multi-mode fiber \textbf{b)} Single-mode Fiber.
 We attribute the slight asymmetry in (b) to a non-perfect cleave.
 }
\label{fig:fluorescence}
\end{figure}

The functionality of the fiber is maintained during the sealing process.
Collecting fluorescence light originating from a free space propagating beam
(c.f.\,Fig.\,\ref{fig:plots}(b))
works as well as monitoring the fluorescence originating from light coming out of the fiber.
Fig.\,\ref{fig:fluorescence} shows the characteristic cone of fluorescing excited atoms around the fiber tip, 
emerging from the numerical aperture.
Transmission measurements with the fibers before and after the sealing process
show no change in transmission performance.
Also, there is no leakage light visible in Fig.\,\ref{fig:fluorescence}.
However,
the mechanical and thermal stress the fiber is exposed to
might lead to changes in polarization dependent properties.

\section{Summary/Outlook}
We described a fabrication method for fiber feedthroughs,
which is both high vacuum and alkali vapor compatible.
The performance of such devices was tested using helium leak checking,
spectroscopic analysis and fluorescence monitoring.

It is now possible to attach the feedthrough to conventional Conflat vacuum flanges
to facilitate a bakeable fiber access to generic vacuum chambers,
or to directly interface it to pump free vapor cells.
Especially sensing applications for electromagnetic fields greatly benefit from the technology,
because the use of metal was completely avoided and only dielectric materials are involved.
Spectroscopy of alkali atoms inside hollow core fibers \cite{rfdressing} in particular provide a promising starting point for such a sensor,
and becomes feasible with the presented technique.

\textbf{Funding.}
Baden-W\"urttemberg Stiftung within the project Micro2Sens, BMBF within Q.Com-Q (Project No. 16KIS0129).
\textbf{Acknowledgment.}
Arzu Yilmaz acknowledges funding by the Deutsche Telekom Stiftung.
Harald K\"ubler acknowledges personal support from the Carl-Zeiss foundation.
The authors thank Frank Schreiber and Joachim Lef\`{e}vre for technical support, Mario Hentschel for the SEM pictures and the Max Planck Institute for the Science of Light for providing the HC-PCF.

\bibliography{references}

\end{document}